\documentclass[
reprint,
superscriptaddress,
amsmath,amssymb,
aps,
prx,
longbibliography]{revtex4-2}

\usepackage{graphicx}
\usepackage{natbib}
\usepackage{amsmath}
\usepackage{amssymb}
\usepackage{appendix}
\usepackage{comment}
\usepackage{relsize}
\usepackage[bottom]{footmisc}
\usepackage[dvipsnames]{xcolor}
\usepackage[section]{placeins}
\usepackage{layouts}
\usepackage{bbold}
\definecolor{darkblue}{rgb}{0,0,.65}
\definecolor{darkgreen}{rgb}{0.28,0.41,0.19}

\usepackage[%
    pdfauthor={David J. Luitz},%
  pdfstartview=FitH,%
  breaklinks=true,%
  bookmarks=true,%
  colorlinks=true,%
  anchorcolor=black,%
  citecolor=blue,
  filecolor=black,%
  menucolor=black,%
  urlcolor=darkblue,%
  linkcolor=blue,%
 ]{hyperref}
\usepackage[all]{hypcap}
\usepackage{braket}

\graphicspath{{./figures_cropped/}}

\begin{document}

\title{
{Sub-diffusive Thouless time scaling in the Anderson model on random regular graphs}
 }%

\author{Luis Colmenarez}
\email{lcolmena@pks.mpg.de}
\affiliation{Max Planck Institute for the Physics of Complex Systems, N\"othnitzer Stra{\ss}e~38, 01187-Dresden, Germany}
\author{David J. Luitz}
\email{david.luitz@uni-bonn.de}
\affiliation{Physikalisches Institut, Universit\"at Bonn, Nussallee 12, 53115 Bonn, Germany}
\affiliation{Max Planck Institute for the Physics of Complex Systems, N\"othnitzer Stra{\ss}e~38, 01187-Dresden, Germany}
\author{Ivan M. Khaymovich}
\email{hai@pks.mpg.de}
\affiliation{Max Planck Institute for the Physics of Complex Systems, N\"othnitzer Stra{\ss}e~38, 01187-Dresden, Germany}
\affiliation{Institute for Physics of Microstructures, Russian Academy of Sciences, 603950 Nizhny Novgorod, GSP-105, Russia}
\author{Giuseppe De Tomasi}
\email{detomasi@illinois.edu}
\affiliation{Department of Physics, University of Illinois at Urbana-Champaign, Urbana, Illinois 61801-3080, USA}
\date{\today}%

\begin{abstract}
The scaling of the Thouless time with system size is of fundamental importance to characterize dynamical properties in quantum systems. In this work, we study the scaling of the Thouless time in the Anderson model on random regular graphs with on-site disorder. We determine the Thouless time from two main quantities: the spectral form factor and the power spectrum. Both quantities probe the long-range spectral correlations in the system and allow us to determine the Thouless time as the time scale after which the system is well described by random matrix theory. We find that the scaling of the Thouless time is consistent with the existence of a sub-diffusive regime anticipating the localized phase. Furthermore, to reduce finite-size effects, we break energy conservation by introducing a Floquet version of the model and show that it hosts a similar sub-diffusive regime.
\end{abstract}

\maketitle

\section{Introduction}
Understanding the emergence of ergodicity in closed quantum many-body systems has received a lot of attention in the last two decades~\cite{Deutsch1991quantum,Srednicki1994chaos,Srednicki1996thermal,rigol2008thermalization,Polkovnikov2011colloquium,DAlessio2016from}. Despite the unitary dynamics, it is believed that generic quantum many-body systems locally thermalize under their dynamics. However, in a seminal work, Basko, Aleiner, and Altshuler~\cite{Basko2006metal} have provided evidence that interacting systems subject to quenched disorder can undergo a transition separating an ergodic/thermal phase from a localized one even at finite temperature.
This phenomenon generalizes the paradigm of Anderson localization~\cite{Anderson1958Absence} to the case of interacting particles and has stimulated extensive research on the resulting many-body localization (MBL).
Subsequently, several works have confirmed the existence of the transition numerically~\cite{oganesyan2007localization,znidaric2008many-body,Pal2010many,Luitz2015many,Corps2021Signatures} and characterised different aspects of the two phases, ranging from dynamical entanglement production to Fock-space structure~\cite{znidaric2008many-body,Bardarson2012unbounded,Serbyn2013universal, Bera2015many, DeTomasi2017quantum,Serbyn2016power, Serbyn2014interferometric,Vasseur2015quantum,Gopalakrishnan2016regimes,Gopalakrishnan2020dynamics,Tikhonov2018many,Mace2019multifractal,luitz2019multifractality,Tarzia2020many,DeTomasi2021rare,Roy2019percolation,Roy2019exact,Roy2020fock}. In the ergodic phase, the system locally thermalizes and the eigenstate thermalization hypothesis holds \cite{Pal2010many,oganesyan2007localization, Deutsch1991quantum,Srednicki1994chaos,Srednicki1996thermal,rigol2008thermalization,Polkovnikov2011colloquium,luitz2016long,Luitz2016anomalous,DAlessio2016from,roy2018anomalous,colmenarez2019statistics}.
The MBL phase is characterized by the emergence of a robust form of integrability, which is described by an extensive number of {\it quasi-local} integrals of motion~\cite{serbyn2013local, huse2014phenomenology,ros2015integrals,Imbrie2016many}. The existence of an MBL phase has opened new exciting possibilities, e. g., the discovery of new quantum phases of matter out-of-equilibrium such as discrete time crystals~\cite{Lazarides2016phase,Lazarides2015fate, Wilczek2012quantum,Sacha2015modeling,Else2016floquet,Yao2017discrete} and might have important applications to quantum memory realizations.
\begin{figure}
    \centering
    \includegraphics[width=0.75\columnwidth]{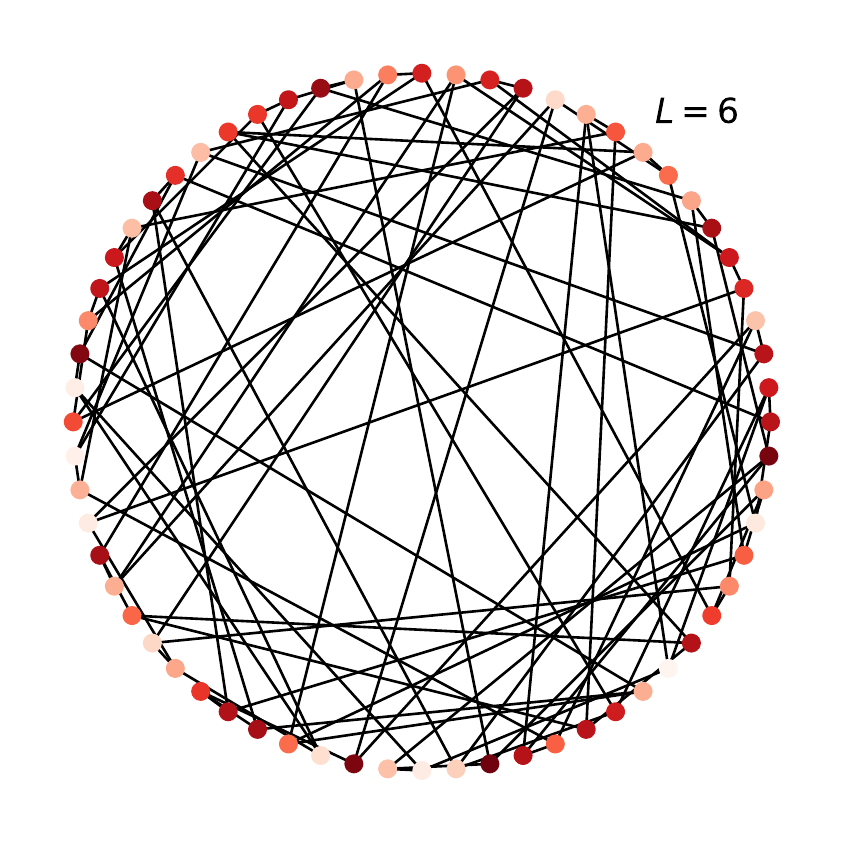}
    \caption{Sketch of a random regular graph (RRG) with diameter $L=6$ and $N=2^L=64$ nodes and branching number $K=2$, i.e. each node has $K+1=3$ neighbors. The particle can hop along the edges (black lines) of the graph with hopping amplitude $t=1$ and is subject to on-site disorder $\mu_i$ (color intensity) on the nodes of the graph (red dots of different intensity).}
    \label{fig:graph}
\end{figure}

Although MBL has been extensively studied, many of its aspects remain puzzling and they are still under intense debate. For instance, the mechanism and nature of the transition is still unclear (see also the discussion in \cite{Morningstar2021avalanches}). This is for instance exemplified by early scaling attempts which yield critical exponents in conflict with general constraints, i.e., so-called Harris bounds~\cite{Goremykina2019analytically,Dumitrescu2019kosterlitz,Morningstar2019renormalization,Morningstar2020many, Laflorencie2020chain, Chandran2015finite, DeRoeck2018many,Luitz2017how,DeRoeck2017many}.
The Fock-space structure and the multifractal nature of the
eigenstates close to the MBL transition within the ergodic side have been at the center of a recent spur of research activity~\cite{Mace2019multifractal,luitz2019multifractality,Tarzia2020many,DeTomasi2021rare,Roy2019percolation,Roy2020fock}.
In the ergodic phase, numerical simulation  has found sub-diffusive transport at intermediate time scales~\cite{BarLev2015absence,Griffiths2015anomalous,Luitz2016extended,Luitz2016anomalous,Khait2016spin,Znidaric2016Diffusive,BarLev2017transport,Bera2017density,agarwal2017rare,luitz2017ergodic,Lezama2019apparent,roy2018anomalous}, contrary to the expected diffusive behaviors of a metal. The sub-diffusion has been argued to stem from the existence of rare Griffith regions which suppress transport. However, the existence of these regions and of the corresponding sub-diffusive behavior as a phase, separated from the diffusive one, is still under intense investigation and it is not clear whether a phase transition between diffusion and sub-diffusion exists. Furthermore, several recent works have been questioning the existence of a genuine MBL transition~\cite{Suntajs2020quantum,Sels2021dynamical,Kiefer-Emanouilidis2020slow}. In these works, using exact diagonalization (ED), a systematic flow towards the ergodic regime has been observed with increasing system sizes, which would imply that MBL is not a stable phase of matter. Subsequent works have pointed out however that this result might be due to the limitation in system sizes reached using ED~\cite{Abanin2021distinguishing,panda2020can,Sierant2020thouless}, or by a large finite-size crossover regime before the MBL phase~\cite{Luitz2020absence,Morningstar2021avalanches,Sels2021markovian}.
For instance, also systems having a firmly established metal-insulator transition, such as the Anderson model on random regular graphs (RRGs) (locally tree-like graphs without boundaries, see Fig.~\ref{fig:graph})~\cite{Abanin2021distinguishing,Sierant2020thouless}, present similar finite-size corrections compared to the MBL problem in one dimension.

Settling this controversy is hindered by the exponential scaling of the Hilbert space with the size of the system and other approaches to make progress are therefore needed.
Given these difficulties, a promising route is to focus on more tractable models, that reproduce the main features of MBL systems. Following the original idea of mapping a disordered quantum dot to a localization problem in the Fock space~\cite{Altshuler1997quasiparticle}, the problem of Anderson localization in hierarchical structures~\cite{AbouChacra1973self}, such as RRGs, has been suggested to be a useful proxy to describe MBL systems~\cite{Biroli2012difference,DeLuca2014anderson}. In this spirit, the sites of the RRG are interpreted as Fock space basis states of the non-interacting model, the on-site energies as renormalized random fields/potentials, and the hopping between sites as an interaction connecting non-interacting states.
The reduction of the (one dimensional) MBL problem to the Anderson model on the RRG is a simplification on several accounts:
(i)~the structure of the Fock space is a hypercube with extensive connectivity of nodes, while in RRG models the connectivity is fixed and intensive; (ii)~the disorder in real space translates to correlated disorder in Fock space, while in the RRG model the on-site energies are uncorrelated; (iii)~the hypercube has many short-range loops unlike RRG, but for both structures the typical loop size is a finite fraction of the graph diameter.

Despite these differences, a range of works have shown that the two models share similarities in their ergodicity-breaking behavior~\cite{Altshuler2016nonergodic,Altshuler2016multifractal,Kravtsov2018nonergodic,Garcia-Mata2017scaling,Parisi2019anderson,Kravtsov2020localization,Tikhonov2016fractality,Tikhonov2017multifractality,Gorsky2020localization,Gorsky2021interacting,Garcia-Mata2020two},
dynamical~\cite{Biroli2017delocalized,Bera2018return,DeTomasi2019subdiffusion,Biroli2020anomalous,Tikhonov2019statistics,Tikhonov2021eigenstate,Khaymovich2021dynamical} properties, as well as finite-size corrections~\cite{Tikhonov2016anderson,Tikhonov2021from,Biroli2018delocalization,Khaymovich2020fragile}.
Thus, a better understanding of the Anderson model on the RRG could shed light on the MBL problem.

Furthermore, the Anderson model on the RRG is not just a proxy for the interacting problem but exhibits a rich and interesting phenomenology on its own. Recently, it has been argued that the Anderson model on the RRG might host a new intermediate phase in between the ergodic and the localized one. This phase, dubbed non-ergodic extended (NEE) phase~\cite{Biroli2012difference,DeLuca2014anderson,Altshuler2016nonergodic,Altshuler2016multifractal,Kravtsov2018nonergodic}, might be composed by critical/multifractal states, i. e., states having strong space fluctuations.
Thus, unlike the Anderson model on the hypercubic lattice $\mathbb{Z}^{d}$ with $d>2$, where multifractal states appear only at the localization transition point, in the RRG an entire phase composed by critical states might exist~\footnote{This discovery has been also found in other random-matrix ensembles~\cite{Kravtsov2015random,Roy2018multifractality,Monthus2017multifractality,Nosov2019correlation,Nosov2019robustness,Biroli2021levy,Kutlin2020renormalization, Kutlin2021emergent,Kravtsov2020localization,Khaymovich2020fragile,Khaymovich2021dynamical,Nosov2021statistics,kutlin2023anatomy}.}.
However, several works have pointed out that this intermediate phase in RRG might be  merely a finite-size effect and therefore ergodicity would be restored in thermodynamic limit ~\cite{Tikhonov2016anderson,Tikhonov2017multifractality,Tikhonov2019statistics,Tikhonov2021from,Khaymovich2020fragile,Khaymovich2021dynamical}.
Another intriguing suggestion is the possible existence of a sub-diffusive phase. By inspecting the spread of a particle initially localized in one of the sites of the RRG, Refs.~\onlinecite{Biroli2017delocalized,Bera2018return,DeTomasi2019subdiffusion,Biroli2020anomalous} have provided evidence that the transport at finite time scales could be sub-diffusive. This sub-diffusive propagation has been found for a range of disorder strengths within the extended phase and long-time scales. This propagation should be compared with the behavior of the Anderson model on the hypercubic $Z^d$ lattice, where sub-diffusion happens only at the critical point.

\begin{figure}[t!]
	\centering
	\includegraphics[width=0.8\columnwidth]{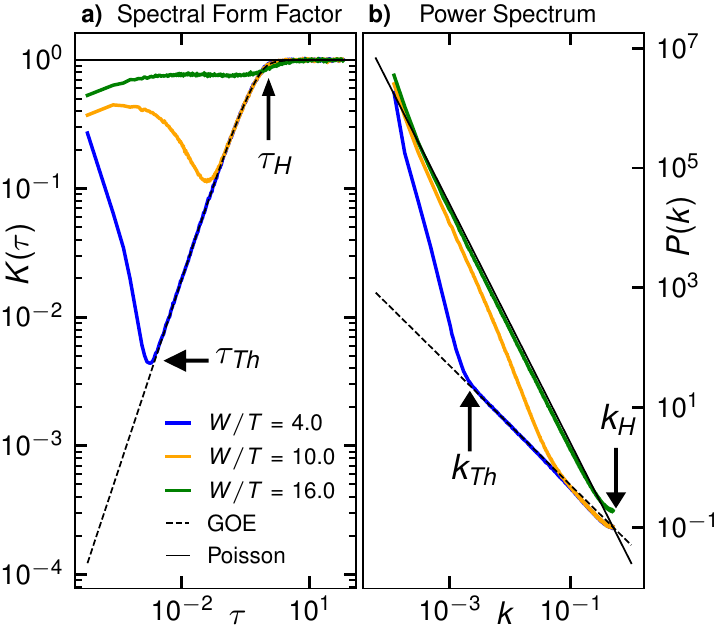}
	\caption{Spectral form factor (a) and power spectrum (b) of Floquet model Eq.~\eqref{eq:model_floquet} for the system size $L=14$ and disorder strength $W/T=4.0,10.0,16.0$ in units of the Floquet half-period $T$. The Thouless time $\tau_\text{Th}$ and Thouless momentum $k_\text{Th}$ are defined as the point where the spectral form factor or power spectrum fit aligned with the GOE prediction. The black dashed lines are the GOE behavior, $K_\text{GOE}(\tau)\sim \tau$ and $P_\text{GOE}(k)\sim 1/k$. The black solid lines are the Poisson behavior, $K_\text{Poisson}(\tau)\sim O(1)$ and $P_\text{Poisson}(k)\sim 1/k^2$.
	The Heisenberg time is the point where the GOE and Poisson values meet at $\tau=1$ and $k=0.5$.}
	\label{fig:figure1}
\end{figure}

Our work aims to shed light on the existence of sub-diffusion in the Anderson model on the RRG solely from the spectral perspective.
Our main probes are the spectral form factor and the power spectrum.
Both, the spectral form factor~\cite{Suntajs2020quantum,Sierant2020thouless,Abanin2021distinguishing} and the power spectrum~\cite{Riser2017power, Riser2020nonperturbative, Riser2021power, Berkovits2020super, Berkovits2021probing} are measures of long-range correlations between eigenvalues of the Hamiltonian. Importantly, these two measures are efficient probes for the dynamics in the system. The Thouless time, $t_{\text{Th}}$, defined as a time scale beyond which the system dynamics is universal and described by the random matrix theory of Gaussian ensembles~\cite{Mehta2004random} (see Fig.~\ref{fig:figure1}) can be extracted from the spectral form factor and the power spectrum. The Thouless time can be interpreted as the time that a particle needs to diffuse throughout the system. As a result, depending on the scaling of $t_{\text{Th}}\sim L^{1/\beta}$ with system size $L$ (diameter of the graph in the case of RRG), one can define different kinds of propagation based on the scaling exponent $\beta$, ranging from super-diffusion, $\beta>1$, and diffusion, $\beta=1$ to sub-diffusion, $0<\beta<1$, and localization, $\beta\to0$. In particular, the Thouless time should be compared to the Heisenberg time, $t_{\text{H}}\sim 2^{L}$, which is the largest meaningful time scale and is defined as the inverse level spacing of the energy spectrum. The system is localized if $t_{\text{Th}}/t_{\text{H}}$ goes to a finite constant in thermodynamic limit. By analyzing the scaling of $t_{\text{Th}}$ with system size $L$, we confirm the existence of a sub-diffusive regime and discuss its finite-size effects. To better pin down the sub-diffusive behavior by reducing finite-size effects, we allow energy fluctuations by introducing a Floquet version of the Anderson model on the RRG. In agreement with the static model, we find also for the Floquet model sub-diffusive transport for a wide range of parameters anticipating the Anderson transition.

The rest of the work is organized as follows. In Sec.~\ref{sec:model} we define the static and Floquet models and discuss their phase diagrams. In Sec.~\ref{sec:Methods}, we introduce the two dynamical probes, the spectral form factor, and the power spectrum. The main results of our work are presented in Sec.~\ref{sec:results} and Sec.~\ref{sec:conclusion} contains concluding remarks.

\section{Models}\label{sec:model}
\begin{figure}[t]
	\centering
	\includegraphics[width=0.8\columnwidth]{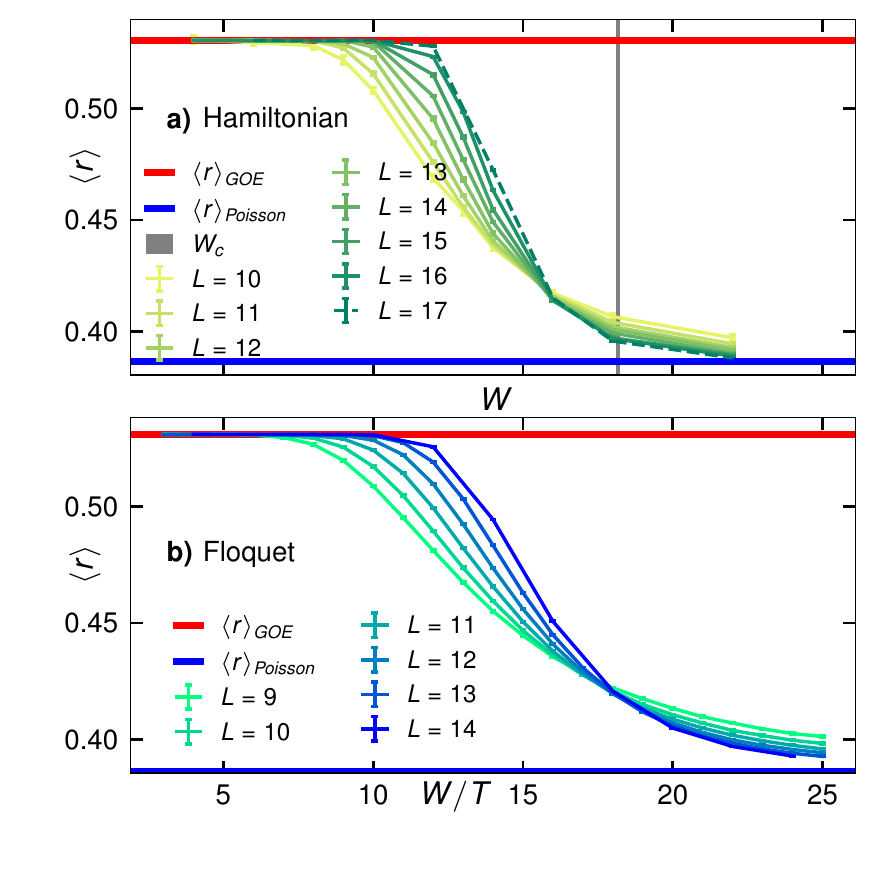}
	\caption{Average consecutive level-spacing ratio $\langle r \rangle$ as a function of disorder $W$ ($W/T$) for different matrix size $N=2^L$. (a)~$\langle r \rangle$ of 10 \% of the Hamiltonian spectrum centered at energy $E=0$ using $2000-5000$ disorder realizations for sizes $L<16$ and $\leq 1000$ for $L=17$. The grey line is the critical disorder $W_c=18.17\pm 0.01$ \cite{Tikhonov2021from}. (b)~$\langle r \rangle$ for Floquet model Eq.~\eqref{eq:model_floquet} using $10000$ disorder realizations for $L<13$ and $4000-6000$ for $L=14$. The solid red and blue lines are the GOE and Poisson value, respectively. Error bars are given by $68 \%$ bootstrap confidence interval.}
	\label{fig:gap_ratio}
\end{figure}

We consider a single particle hopping on a RRG with $N$ sites, subject to on-site disorder $\{\mu_i\}$. The Hamiltonian is given by
\begin{eqnarray}
\label{eq:model}
H = \mathlarger{\sum}_{i=1}^{N} \mu_i|i\rangle\langle i|-\mathlarger{\sum}_{ \{i,j\} \in E }|i\rangle\langle j|,
\end{eqnarray}
where $\{\mu_i\}$ are independent random variables with box distribution $[-W/2,W/2]$ and the sum in the second term runs over the edges $\{i,j\}$ of the graph, i.e. the set of connections between sites in a given realization of the RRG (cf. Fig.~\ref{fig:graph} for an example, where edges are indicated by lines).
Each site of an RRG has a fixed number of neighbors $K+1$, where $K$ is the branching number, while the precise configuration of edges in the graph is subject to random sampling. In this work, we focus on the smallest nontrivial branching number $K=2$ different from a one-dimensional problem, such that each site has $3$ neighbors.
The total number of nodes in the graph is $N=2^L$ but we refer to it via the diameter of the graph $L\sim \ln N$ in order to make the analogy with the Hilbert space of $1/2$-spin chains of length $L$.
This model has an Anderson localization transition at $W_{AT}\approx 18.1$~\cite{DeLuca2014anderson,Kravtsov2018nonergodic,Parisi2019anderson,Tikhonov2021from}, which is blurred by a finite size crossover regime and appears at lower disorder in finite graphs ~\cite{Tikhonov2016anderson,Tikhonov2017multifractality,Tikhonov2019statistics,Tikhonov2021from,Khaymovich2020fragile,Khaymovich2021dynamical, Morningstar2021avalanches}. For instance the average level spacing for $L=17$ has crossings close to $W\approx16$ (see Fig.~\ref{fig:gap_ratio}).

The Hamiltonian in Eq.~\eqref{eq:model} hosts a single-particle mobility edge, which separates extended from localized states as a function of energy and its density of states has a non-trivial shape peaked close to the edges of the spectrum (see Appendix~\ref{appendix:unfolding} and cf.~\cite{Valba2021mobility}).
In order to overcome finite-size effects from the localized energy bands and from the energy dependence of the density of states, we introduce a Floquet version of the Anderson model on RRG
\begin{eqnarray}
\label{eq:model_floquet}
U = \exp(-\mathrm{i} H_1 \, T)\exp(-\mathrm{i} H_2 \, T),
\end{eqnarray}
where $H_1$ and $H_2$ are hopping Hamiltonians as in Eq.~\eqref{eq:model} on the \emph{same} RRG but with different diagonal disorder $\{\mu_i\}$. $2T$ is the driving period and we set $T=2$ for the rest of the work. The results are not qualitatively affected by the concrete value of $T$, see Appendix~\ref{appendix:different_periods} for details. Both models, static and periodic, have transitions from Wigner-Dyson to Poisson level statistics as can be seen in the average gap ratio $r$ in Fig.~\ref{fig:gap_ratio}.

The Anderson localization transition in the static case, Eq.~\eqref{eq:model}, has been extensively studied~\cite{AbouChacra1973self, Biroli2012difference, DeLuca2014anderson, Altshuler2016nonergodic, Altshuler2016multifractal, Kravtsov2018nonergodic, Parisi2019anderson, Tikhonov2016fractality, Tikhonov2017multifractality, Gorsky2020localization,Gorsky2021interacting, Biroli2020anomalous,Tikhonov2019statistics,Tikhonov2021eigenstate,Khaymovich2021dynamical, Tikhonov2016anderson,Tikhonov2021from,Biroli2018delocalization,Khaymovich2020fragile}
and we expect the transition in the driven model to have similar features. The advantage of introducing a Floquet drive is the flat density of states on the unit circle, which allows us to perform an unbiased study of long-range spectral correlations.

In the following we refer the eigenvalues $\{E_n\}$ of the static model,  Eq.~\eqref{eq:model}, as ``spectrum'' and  $\{\theta_n = \mathrm{arg}(\omega_n)\}$ of the driven model, Eq.~\eqref{eq:model_floquet}, as the eigenphases or quasienergies, where $\{\omega_n=e^{i\theta_n}\}$ are the eigenvalues of $U$, which lie on the complex unit circle. We expect that the Floquet model~\eqref{eq:model_floquet} yields cleaner results for correlations in the spectrum at large phase differences between the eigenvalues, which is confirmed by our numerical results.

\section{Methods}\label{sec:Methods}

In this section, we introduce the quantities used to study spectral correlations and detail the computation of the Thouless time.

\subsection{Level spacing}
We start with a well known probe to detect short range spectral correlations, which are captured by the statistics of spacings between two adjacent eigenvalues
\begin{eqnarray}
\label{eq:level_spacing_definition}
s_i := E_{i+1}-E_{i}.
\end{eqnarray}

The spacing ratio, defined as
\begin{eqnarray}
\label{eq:gap_ratio_definition}
r_i = \dfrac{\min(s_i,s_{i+1})}{\max(s_i,s_{i+1})} \ ,
\end{eqnarray}
has been found to be a useful resource to separate a delocalized phase from a localized one~\cite{oganesyan2007localization,Atas2013distribution}. In an ergodic phase, the spacings between eigenvalues are distributed like the ones of a Gaussian random matrix. In contrast, in a localized phase energy levels tend to cross each other as a function of a parameter with little interaction between eigenvalues and the spacings are thus Poisson distributed. In particular, for the Gaussian orthogonal ensemble (GOE) the average spacing ratio is given by the Wigner-Dyson value $\langle r \rangle_{GOE} \approx 0.53590 $, while for Poisson spectra it is $\langle r \rangle_{Poisson} =2\ln 2 -1 \approx 0.38629$~\cite{oganesyan2007localization,Atas2013distribution} where $\langle...\rangle$ indicates averaging over disorder realization of graph and diagonal disorder as well as eigenstates.

The averaged energy spacing is the smallest energy scale in the problem and it defines the Heisenberg time,
\begin{equation}\label{eq:Heisenberg_tim}
t_{\text{H}}=2\pi/\langle s_i\rangle.
\end{equation}
Consequently, the Heisenberg time $t_\text{H}$ is the largest meaningful time scale in a finite system, on which discrete energy levels can be resolved. It is proportional to the dimension of the Hilbert space $t_H\propto 2^L$.

\subsection{Spectral unfolding}\label{sec:unfolding}

The $r$-statistics defined in Eq.~\eqref{eq:gap_ratio_definition} provides only limited insight into the dynamics of the system because it only involves the computation of small energy scales (asymptotically large times).
As a result, the $r$-statistics can separate only a delocalized phase from a localized one, without providing information about the  transport properties.

To inspect finite time scales relevant for the transport, we have to consider the long-range spectral correlations. However, before defining the two main measures to analyze energy correlation, we first need to introduce the notion of spectrum unfolding. To remove the influence of the non-uniform density of states in the spectra of the static Hamiltonian, Eq.~\eqref{eq:model}, we perform an unfolding of the spectrum, which maps the eigenvalues $\{E_n\}$ of the Hamiltonian with non-uniform density of states  to the ``unfolded'' eigenvalues $\{\varepsilon_n\}$, which have a homogeneous density of states.

Instead of working directly with the density of states, we use the cumulative distribution function (CDF), which is defined by the fraction of eigenvalues smaller than $E$:
\begin{equation}
    \mathrm{CDF}(E) = \frac{\#(E_n < E)}{N},
\end{equation}
where for a given energy $E$, $\#(E_n<E) \in \mathbb{N}$ is the number of eigenvalues smaller than $E$.
For each disorder realization, this function is the empirical CDF, a step-wise function with steps of size $1/N$ at the positions of the eigenvalues $\{E_n\}$.

In order to obtain the average CDF over $n_\text{real}$ disorder realizations, we combine the spectra $\{E_n^{(i)} \}$ of all realizations and sort the resulting $n_\text{real} N$ values. The empirical CDF of these joint spectra is then a step function with steps of size $1/(n_\text{real}N)$ at the positions of the sorted values $\{E_n^{(i)}\}$.
For practical purposes, the obtained average empirical integrated density of states $\text{CDF}^\text{avg}(E)$ is smoothed by a cubic spline to minimize statistical fluctuations.

Each eigenvalue $E_n^{(i)}$ of a realization $(i)$ can then be mapped to its unfolded and normalized counterpart
\begin{equation}
    \label{eq:unfolded_spectrum}
    \varepsilon_n^{(i)}/N = \text{CDF}^\text{avg}(E_n^{(i)}) \in [0,1],
\end{equation}
where the density of states of the unfolded $\varepsilon_n^{(i)}$ is constant. The procedure is illustrated in Fig.~\ref{fig:unfolding} in Appendix~\ref{appendix:unfolding}. For measuring the Thouless time in units of the mean level spacing, it is more convenient to work with an equidistant spectrum with unit mean level spacing, thus we work with the unfolded spectrum:
\begin{equation}
    \label{eq:unfolded_spectrum_unit_mls}
    \varepsilon_n^{(i)} \in [0,N] \ .
\end{equation}
Note that in the case of the Floquet model defined in Eq.~\eqref{eq:model_floquet}, spectral unfolding is not needed and the simple rescaling $\varepsilon_n^{(i)}= N \theta_n^{(i)}/2\pi$ yields a uniform spectrum with unit level spacing similar to $\varepsilon$ in Eq.~\eqref{eq:unfolded_spectrum_unit_mls}. Henceforth the realization index $(i)$ will be dropped from the unfolded spectrum $\{\varepsilon^{(i)}_{n}\}$ and we use only $\{\varepsilon_{n}\}$ to denote the $n$-th level of the spectrum, making the disorder index implicit.

\subsection{Spectral form factor}\label{sec:model_sff}

The first probe that we introduce is the spectral form factor, defined as~\cite{Mehta2004random}
\begin{eqnarray}
\label{eq:sff_definiton}
	K(\tau) =\left\langle \dfrac{1}{\sum_{n}g^2(\varepsilon_n)}\Big |\mathlarger{\sum}_{n}  g(\varepsilon_n)e^{-i2\pi\varepsilon_n\tau}\Big |^2
	\right\rangle \ ,
\end{eqnarray}
where $\{\varepsilon_n\}$ is the Hamiltonian unfolded spectrum or the Floquet quasienergy spectrum. The function
\begin{equation}
\label{eq:sff_filter}
g(\varepsilon_n) =
\begin{cases}
\exp\left[-(\varepsilon_n-\bar{\varepsilon})^2/2\eta^2\sigma^2 \right], & \text{Hamiltonian} \\
1, & \text{Floquet}
\end{cases}
\end{equation}
is a filter for mitigating spectral-edge effects in the static unfolded spectra. $\bar{\varepsilon}$ is the center of the spectrum for a given disorder realization, $\sigma$ is the width of the spectrum and $\eta$ is a parameter which controls the relative width of the filter. To avoid finite size-effects from the localized energy-bands we focus on only $60\%$ of the states centered at the middle of the spectrum $\bar{\varepsilon}$. In addition, we apply the Gaussian filter mentioned above and set $\eta=0.3$ through the entire work, except where specified otherwise. The definition of spectral form factor given in Eq.~\eqref{eq:sff_definiton} guarantees $K(\tau)=1$ at asymptotic large times $\tau\approx \tau_H=1$. The spectral form factor for GOE spectra is described by a linear growth, $K_\text{GOE}(\tau)\approx 2\tau-\tau\ln(1+2\tau)$~\cite{Brezin1997spectral} for $\tau<1$, while for a Poisson spectrum, $K_\text{Poisson}(\tau)\equiv 1$.

The Thouless time $\tau_\text{Th}$ identifies the time scale after which the dynamics is described by random matrix theory. In particular, for the times $\tau<\tau_\text{Th}$ the quantum dynamics is governed by the locality of the system, quantum correlations spread dynamically before reaching the boundary of the system. Thus, in the way how $\tau_\text{Th}$ scales with $L$, it is possible to probe different classes of the propagation of correlations, e. g., diffusion, sub-diffusion or localized states. At later times, $\tau>\tau_\text{Th}$, quantum correlations have been scrambled to a large extent across all length scales, and thus at such late times the system becomes indistinguishable from a random matrix model, leading to the GOE behavior of the spectral form factor. Hence, the Thouless time $\tau_\text{Th}$ corresponds to the time beyond which $K(\tau)$ ramps linearly and can be approximated with the behavior of $K_{GOE}(\tau)$, i.e. within a certain threshold ($K(\tau>\tau_\text{Th}) \approx K_\text{GOE}(\tau)$). Figure~\ref{fig:figure1} shows the typical behavior of $K(\tau)$ for a fixed system size and several disorder strengths. As one can observe in Fig.~\ref{fig:figure1}, it is possible to identify a time scale $\tau_\text{Th}$, from which $K(\tau)$ follows the GOE curve (dashed line). In Appendix~\ref{appendix:thouless_time} the procedure for extracting the Thouless time is explained in detail.

To analyze the scaling of the Thouless time with system size, we rescale it with the actual Heisenberg time in Eq.~\eqref{eq:Heisenberg_tim}:

\begin{eqnarray}
\label{eq:thouless_time_definition_sff}
t_\text{Th} = \tau_\text{Th}t_\text{H}.
\end{eqnarray}

As we will discuss later in Sec.~\ref{sec:sub-diffusion_def}, for a diffusive system in a tree-like structure we expect $t_\text{Th}\propto L$ like in the classical diffusion problems on Bethe lattice~\cite{Cassi1989random} or RRG~\cite{Chinta2015heat}, while in the localized phase, the spectrum is Poissonian and thus $K_\text{Poisson}(\tau)\equiv 1$ and $t_\text{Th}=t_\text{H}\propto 2^L$.

\begin{figure*}
	\centering
	\includegraphics[width=0.9\textwidth]{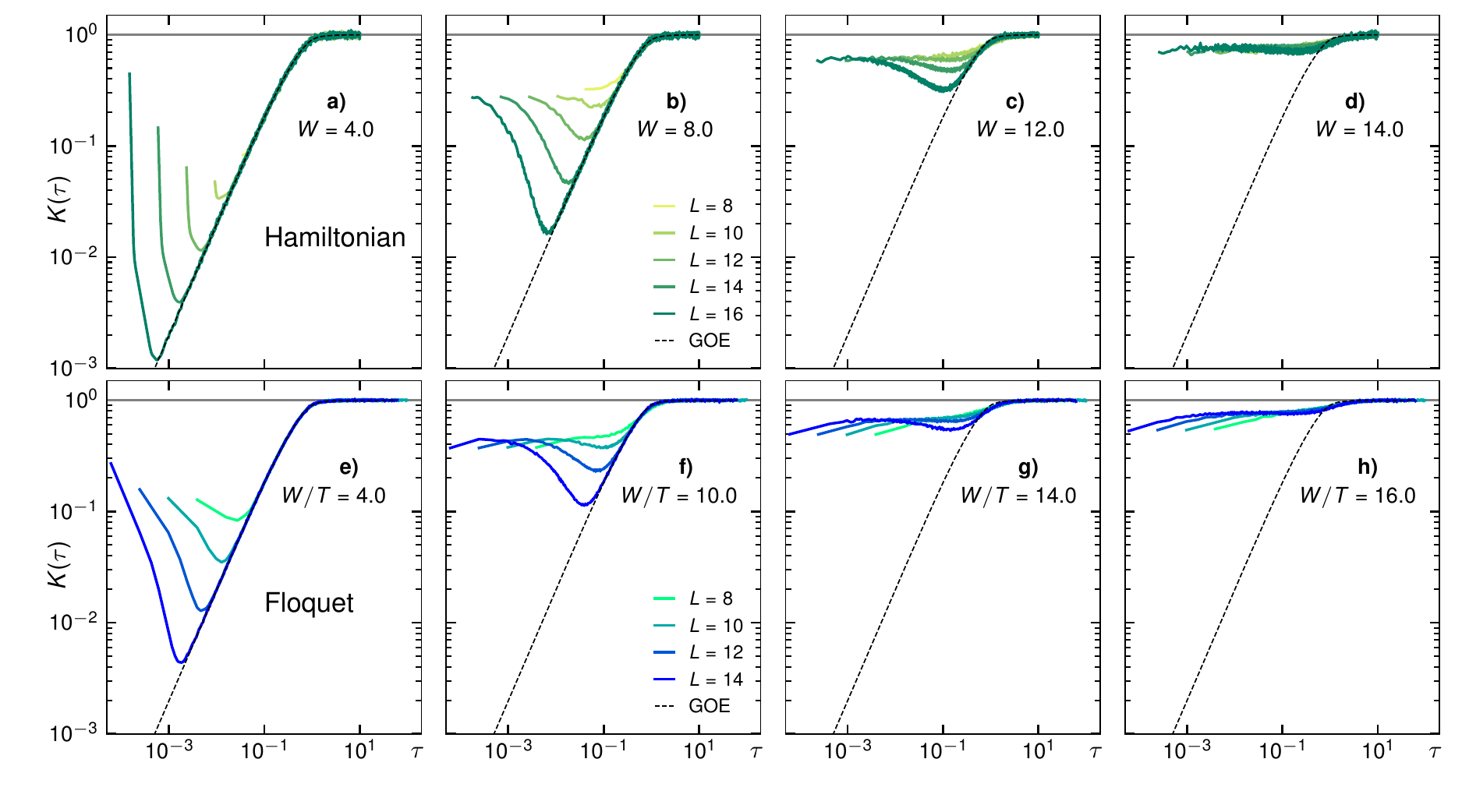}
	\caption{Spectral form factor, Eq.~\eqref{eq:sff_definiton}, for the unfolded Hamiltonian spectrum for different system sizes $L=8$, $10$, $12$, $14$, $16$ at (a)~$W=4.0$, (b)~$8.0$, (c)~$12.0$, (d)~$14.0$  and for Floquet quasienergies for  $L=8$, $10$, $12$, $14$ at (e)~$W/T=4.0$, (f)~$10.0$, (g)~$14.0$, (h)~$16.0$.
	In the static case, only 60\% of the spectrum centered at the middle are taken and the disorder average is performed over 2000-5000 disorder realizations  The time $\tau\in [ 10^{-4},10^{1}]$ is continuous.
	In the Floquet case, time $\tau=t/N$ is discrete and given by an integer $t\in [1,n]$ normalized by the matrix size $N$, with $n=10^6$ for $8<L<14$, $n=10^4$ for $L=8$, and $n=10^7$ for $L=14$.
	}
	\label{fig:sff}
\end{figure*}

\subsection{Power Spectrum}\label{sec:model_ps}
A complementary measure to detect long-range level correlations is the spectral power spectrum. It can be defined through the spectral statistic $\{\delta_n\}$ given by
\begin{eqnarray}
\label{eq:delta_definition}
\delta_n := \varepsilon_n-n,
\end{eqnarray}
with $n\in \{1,...,N\}$ and the Hilbert space dimension $N=2^L$.
For both models, static and Floquet, $\left\langle\varepsilon_n\right\rangle=n$. Consequently, $\{\delta_n\}$ measures the distance of the $n$-th level from a rigid  equidistant spectrum.

The sequence $\{\delta_n\}$ can be interpreted as a discrete ``time'' series with zero mean and the index $n$ as ``time-point''. A useful way to analyze this time series $\{\delta_n\}$ is to consider its Fourier transform, the \emph{power spectrum}:
\begin{eqnarray}
\label{eq:power_spectrum_definition}
P(k)  & = & \left\langle |\mathcal{F}(\delta_n)|^2\right\rangle \nonumber \\ & = & \left\langle\dfrac{1}{\mathcal{N}} \left|\sum_{n=1}^{N} g_n\delta_n\exp\left(\dfrac{-2\pi i kn}{N}\right)\right|^2\right\rangle \ .
\end{eqnarray}
As for the spectral form factor, we have introduced the filter function (cf. Eq.~\eqref{eq:sff_filter})
\begin{eqnarray}
\label{eq:ps_filter}
g(n) =
\begin{cases}
\exp(-(\varepsilon_n-\bar{\varepsilon})^2/2\eta^2 \sigma^2)), & \text{Hamiltonian} \\
1, & \text{Floquet}
\end{cases}, \
\end{eqnarray}
to reduce finite-size energy edge effects for the Hamiltonian model.
$\mathcal{N}=\sum_n g_n^2$ is the normalization constant of the Fourier transform in the filtered spectrum. Like for the spectral form factor, $\bar{\varepsilon}$ and $\sigma^2$ are the mean and variance of the unfolded spectrum for each disorder realization. The strength of the filtering is set to $\eta=0.3$. For reducing edges effects even further, the edges of the spectrum are cut off and only $60\%$ of the spectrum centered at
the middle of the spectrum $\bar{\varepsilon}$.
$1\leq k\leq N_y$ is integer, with $N_y=N/2$ being the largest possible meaningful frequency in the system, called the Nyquist frequency.

For Poisson spectra, when there are no correlations at any range in the spectrum this time series is similar to a sample of a Brownian motion with displacement $\delta_n$. In the limit $k/N\ll 1$ and $N\gg1$, the asymptotic form of the power spectrum is given by $P_\text{Poisson}(k)=1/f^2$, where $f$ is the Fourier frequency $f=2 \pi k/N$~\cite{faleiro2004theoretical}. In the GOE case, the asymptotic form is $P_\text{GOE}(k)=\frac{N}{2\pi k}=1/f$. This $1/f$ noise has been argued to be an unique characterization for quantum chaotic systems~\cite{corps2020thouless,faleiro2004theoretical,relano2002quantum}.

The variable $k$ in the power spectrum does not have units of inverse time, it is rather a dimensionless ``energy momentum'' alluding to the fact that it comes from the argument of a Fourier transform of an energy coordinate. In the same spirit of the spectral form factor, the dimensionless Thouless energy momentum $k_\text{Th}$ is interpreted as the smallest momentum for which $P(k)\sim P_\text{GOE}(k)\propto 1/\tilde{k}$ with $\tilde{k}=k/N$ (see Fig.~\ref{fig:figure1}). $l=1/k$ can be interpreted as an average energy distance in the spectrum, henceforth at distances $l>l_\text{Th}$, with $l_\text{Th} = 1/k_\text{Th}$, the levels are uncorrelated whilst at $l<l_\text{Th}$ they are correlated and well described by random-matrix theory. Figure~\ref{fig:figure1} shows the power spectrum as a function of $k$ for $L=17$ and several disorder strengths. The dashed line in Fig.~\ref{fig:figure1} is the GOE behavior, $P(k)\propto 1/f$, from where one can read the Thouless momentum $k_\text{Th}$.

In analogy with the spectral form factor, the Thouless time is given in units of the Heisenberg time: %
\begin{eqnarray}
\label{eq:thouless_time_definition_ps}
t_\text{Th} = k_\text{Th}t_\text{H} \ ,
\end{eqnarray}
where we use the same expression for $t_\text{H}$ as in Eq.~\eqref{eq:thouless_time_definition_sff}. In Appendix~\ref{appendix:thouless_time} the procedure for computing $k_\text{Th}$ is explained in detail.

\subsection{(Sub-)diffusion and Thouless time}\label{sec:sub-diffusion_def}
Having defined the spectral form factor and the power spectrum, our main probes we will use to detect long-range correlations in the energy spectrum, we now discuss the relations between (sub-) diffusion and the scaling of the Thouless time with $L$.

Let us start by pointing out the differences between the propagation in hypercubic, $\mathbb{Z}^d$-like, lattices and hierarchical structures, e. g., RRG or the Bethe lattice. Diffusive dynamics is usually defined through the spread of an initially localized wave packet. In the $\mathbb{Z}^d$ lattice, diffusion is quantified by the Gaussian profile of a propagating wave-packet. As a result, in the $\mathbb{Z}^d$ lattice the survival probability decays as $\Pi(t)\sim t^{-d/2}$ and the mean squared displacement $\Delta X^2(t) \sim t^{2\beta}$ with $\beta = 1/2$. Sub-diffusive propagation is consequently given by $\beta <1/2$. In particular, the related Thouless time scales with the linear system size as $t_\text{Th} \sim L^{1/\beta}$, which is the time to cross the system. On the other hand, in diffusive processes on tree-like structures the survival probability decays exponentially fast $\sim e^{-\Omega(K)t}$, with a certain decay rate $\Omega(K)$, depending on the branching number $K$, and the mean square displacement grows as $\sim t^{2\beta}$ with $\beta = 1$~\cite{Cassi1989random,Chinta2015heat}. As a result, linear growth of the Thouless time with the diameter $L$ ($t_{Th} \sim L$) implies diffusion and sub-diffusion takes place if $\Delta X^2(t)\sim t^{2\beta}$ ($t_{Th}\sim L^{1/\beta})$ with $\beta < 1$.

In a localized phase, where the degrees of freedom are frozen at long times, the Thouless time is comparable with the Heisenberg time. Indeed, the ratio $t_\text{Th}/t_\text{H}$ can be used as a probe to detect an delocalization-localization transition. In the extended phase $t_\text{Th}/t_\text{H}\rightarrow 0$, while in a localized one $t_\text{Th}/t_\text{H}\sim O(1)$ in thermodynamic limit.

\section{Results}\label{sec:results}
\begin{figure*}
	\centering
	\includegraphics[width=0.9\textwidth]{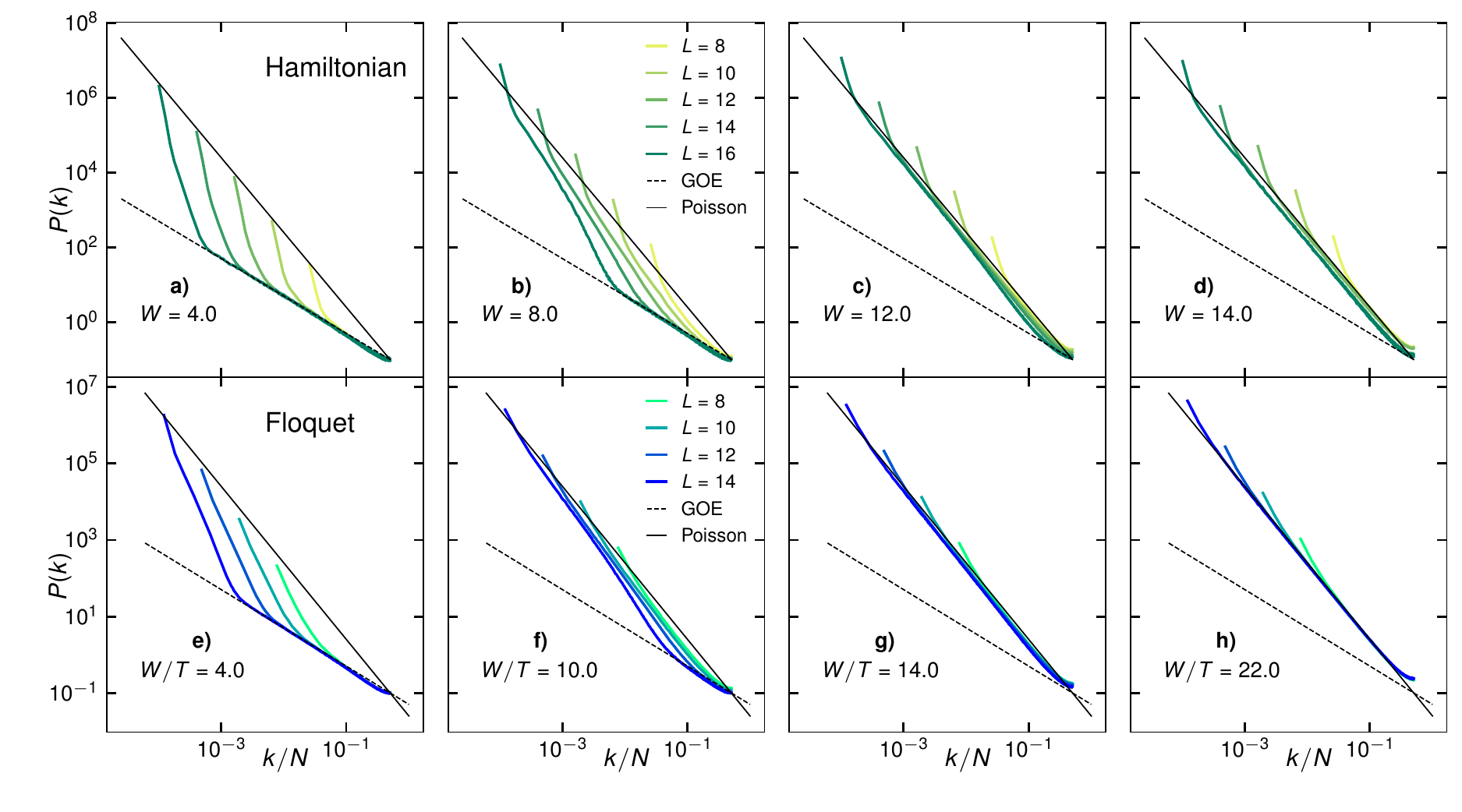}
	\caption{Power spectrum, Eq.~\eqref{eq:power_spectrum_definition},
	for the unfolded Hamiltonian spectrum for different system sizes $L=8$, $10$, $12$, $14$, $16$ at (a)~$W=4.0$, (b)~$8.0$, (c)~$12.0$, (d)~$14.0$  and for Floquet quasienergies for  $L=8$, $10$, $12$, $14$ at (e)~$W/T=4.0$, (f)~$10.0$, (g)~$14.0$, (h)~$22.0$.
	In the static case, only 60 \% of the centered at the middle are taken. The ``energy momentum'' $k$ is normalized by the effective size of the spectrum $\tilde{N} = 0.6 2^L$, thus, the arguments of $P_{GOE}(k)$ and $P_{Poisson}(k)$ are also rescaled by $\tilde{N}$.
	In Floquet case, $k$ is rescaled by the matrix size $N=2^L$ such that the Nyquist frequency becomes $0.5$.}
	\label{fig:ps}
\end{figure*}

In the following, we present results obtained by solving the models in Eqs.~\eqref{eq:model} and~\eqref{eq:model_floquet} using exact diagonalization techniques and computing the spectral form factor and the power spectrum to extract the Thouless time as a function of system size.
Unlike the static Hamiltonian case, the critical disorder, $W_\text{AT}$, for the Anderson transition in the Floquet model, Eq.~\eqref{eq:model_floquet}, is unknown. Thus, before starting our investigation on the long-range correlations, we start with the short-range ones. This preamble will allow us to obtain a rough estimate of the critical value for the Floquet model and, as well as to determine the range of disorder strengths in which finite-size corrections are negligible.

Figure~\ref{fig:gap_ratio} shows the $r$-statistics as a function of disorder strength $W$ and several system sizes $L$ for both models. The critical strength for the static Hamiltonian model in Eq.~\eqref{eq:model} has been estimated analytically at $W_\text{AT}\approx 18.1$~\cite{Kravtsov2018nonergodic,Parisi2019anderson,Tikhonov2021from, DeLuca2014anderson}. As expected at weak disorder the $r$-statistics is GOE and at strong disorder the spectrum is Poisson. For the available system sizes, the $r$-statistics in Eq.~\eqref{eq:gap_ratio_definition} shows crossings close to $W\approx 16 <W_{AT}$ with an apparent drift towards larger $W$ for growing graph diameters $L$, as one can observe in Fig.~\ref{fig:gap_ratio}. Therefore, at $W \gtrsim 16$, when our system sizes are smaller than the correlation volume, the physics will be largely dominated by finite-size effects, and the system looks already localized. Since we are interested in the transport properties of the system, we will hence mainly focus on disorder strengths $W<16$. For our Floquet model of Eq.~\eqref{eq:model_floquet} there is no exact estimate of the critical disorder. The finite-size gap ratio crossover between the ergodic value and the Poisson one happens around $W/T\approx 18$. This value, $W/T\approx 18$, provides us with a lower bound of the exact critical value and we consider disorders $W/T>18$ to be already within the localized phase. A more precise estimate of the critical disorder of the Floquet model is not needed here and beyond the scope of this work. Since we consider transport properties, we focus on the delocalized part of the phase diagram at a safe distance from the critical regime. We therefore restrict the discussion to disorder strengths $W<16$ for the Hamiltonian model and $W/T<18$ for the Floquet model.

\begin{figure*}
	\centering
	\includegraphics[width=0.9\textwidth]{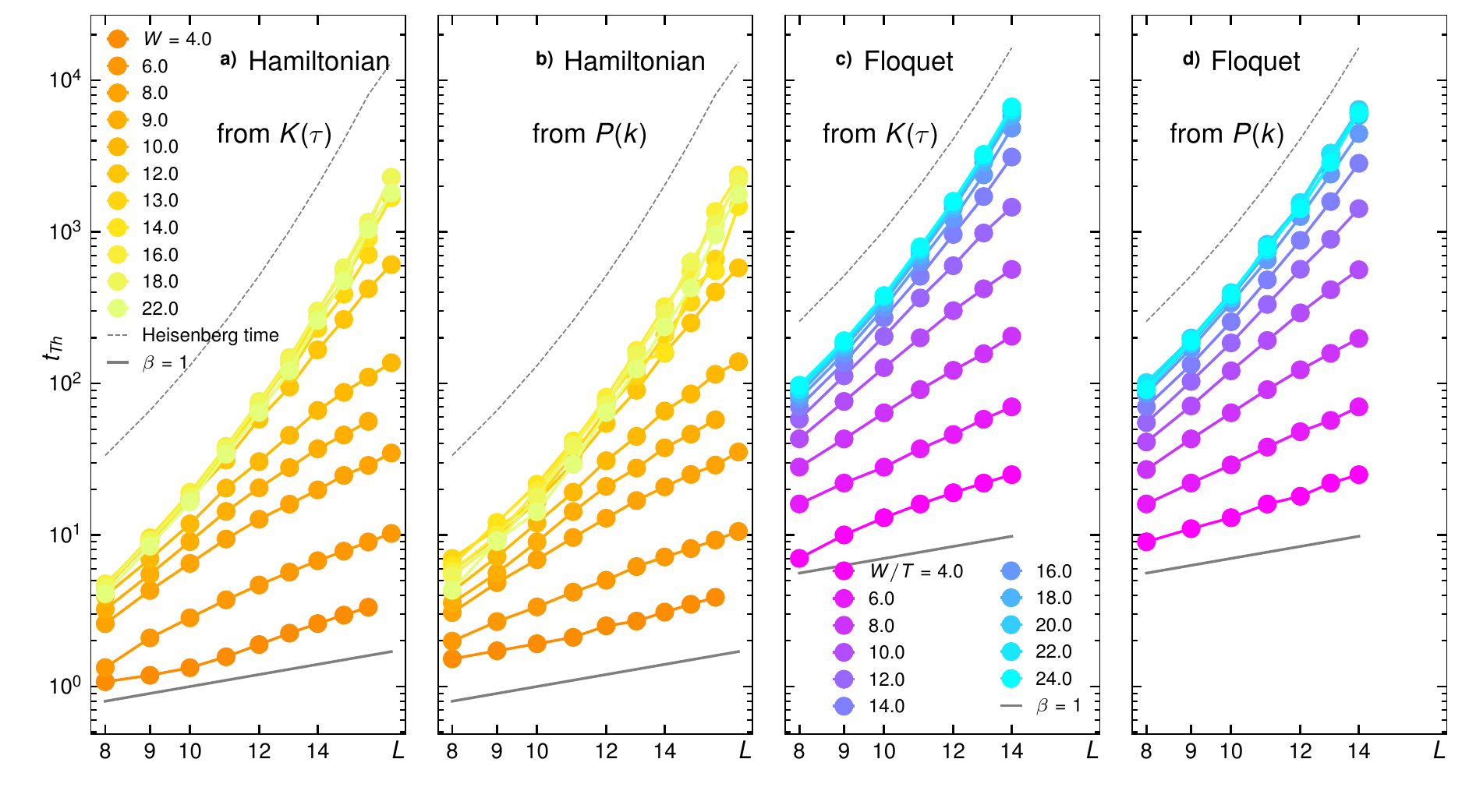}
	\caption{Thouless time extracted from the spectral form factor (a,~c) and the power spectrum (b,~d) as the point, where $\log (K/K_{GOE})=0.4$ or $\log (P/P_{GOE})=0.4$ for Hamiltonian (a,~b) and Floquet (c,~d) cases. Each curve corresponds to the different disorder value shown in the legend. The grey dashed lines show the inverse level spacing (Heisenberg time). The solid grey lines correspond to the diffusive scaling with an arbitrary prefactor. The set of parameters for the computation of the spectral form factor and the power spectrum is the same as the one of Fig.~\ref{fig:sff}  and~\ref{fig:ps}.}
	\label{fig:thouless_time}
\end{figure*}

\begin{figure}
	\centering
	\includegraphics[width=1\columnwidth]{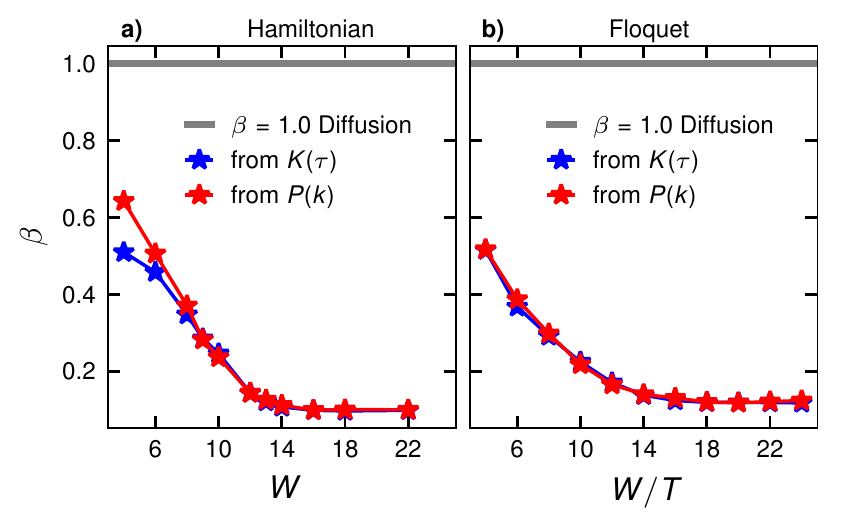}
	\caption{Dynamical exponent $\beta$ for the Hamiltonian~(a) and  Floquet~(b) models, extracted by fitting $t_{Th}\sim L^{1/\beta}$ from the spectral form factor (blue) and the power spectrum (red) considering only the five largest system sizes shown in Fig.~\ref{fig:thouless_time}.
	}
	\label{fig:dynamical_exponent}
\end{figure}

Having ascertained a reliable delocalized regime for our simulations, we first look at the raw data for the spectral form factor $K(\tau)$ and the power spectrum $P(k)$ for the Hamiltonian and Floquet model. The full data, for several system sizes and relevant disorder strengths, can be found in Fig.~\ref{fig:sff} and~\ref{fig:ps}, respectively.
As one can observe in Fig.~\ref{fig:sff}, at weak disorder $W=4$ and $W/T=6.0$, deep in the ergodic phase, the GOE ``ramp'' at late time $\tau$, $K(\tau) \approx 2\tau$, gets longer with the increasing system size.
When approaching the Anderson transition, the GOE ramp gets shorter until it is barely visible at disorder strengths $W=14.0$ and $W/T=16.0$. These disorder amplitudes are not yet within the localized phase according to the mean level spacing in Fig.~\ref{fig:gap_ratio}, although in a finite system they show very slow dynamics signaled by very long Thouless times~\cite{DeTomasi2019subdiffusion}. In Fig.~\ref{fig:ps} we show the power spectrum for the same disorder values. Similarly to $K(\tau)$, at low disorder $W=4.0$ and $W/T=4.0$ the power spectrum falls on top of the GOE prediction ($P(k)\sim 1/k)$, which is shown by the dashed line. Upon increasing disorder the power spectrum meets the GOE curve at larger $k$. At $W=12.0$ and $W/T=14.0$ the GOE scaling region gets shorter but still increases with system size, denoting a flow towards delocalization. On the other hand, at $W=14.0$ and $W/T=16.0$ all accessible system sizes seem to follow the Poisson behavior ($P_\text{Poisson}(k)\sim 1/k^2)$, although at $W=14.0$ the power spectrum has a small visible offset from the Poisson value. This offset is also seen in the Floquet model at $W/T=10.0$ for small system sizes, hence that might be a sign of flow towards delocalization at larger system sizes, compatible with the previously estimated location of the Anderson transition in the static model at $W_\text{AT}=18$.
At very strong disorder, $W/T=22$, where the Floquet model is likely in the localized phase (cf. Fig.~\ref{fig:gap_ratio}), we find excellent agreement of the spectral form factor with Poisson statistics.

Based on this data, we now turn to the main aim of this work, the scaling of the Thouless time with system size $t_\text{Th}\sim L^{1/\beta(W)}$. As discussed in Sec.~\ref{sec:sub-diffusion_def}, the exponent $\beta$ is connected to the transport properties. The Thouless time can be understood as the time for a particle to propagate throughout a graph of diameter $L$. In particular, $\beta = 1$ corresponds to diffusion and $0<\beta <1$ to sub-diffusion.
In Fig.~\ref{fig:thouless_time} we show the Thouless time, $t_\text{Th}$, extracted from $K(\tau)$ and $P(k)$ as a function of $L$ on a doubly logarithmic scale for several disorder strengths $W$. For comparison in Fig.~\ref{fig:thouless_time} the Heisenberg time (dashed line) and the diffusive scaling of the Thouless time with $\beta=1$ (solid grey line) are also reported. The Heisenberg time signals the localization behavior ($t_\text{Th}\sim t_\text{H}$). At weak disorder, the scaling of the Thouless time is indeed compatible with a power law $L^{1/\beta}$, while at strong disorder we observe significant curvature, stemming from the exponential scaling of the Heisenberg time and $t_\text{Th} \sim t_\text{H}$ in the localized phase.

From the apparent power law scaling of the Thouless time $t_\text{Th} \sim L^{1/\beta}$, we extract the dynamical exponent $\beta$, which is reported in Fig.~\ref{fig:dynamical_exponent}. We find that even at weak disorder, in both models $\beta<1$, corresponding to sub-diffusion propagation~\cite{DeTomasi2019subdiffusion}. Indeed, for disorders $W=6.0$ and $W/T=6.0$ the dynamical exponent $\beta\approx 0.5$ and, then decreases upon increasing disorder as expected. At stronger disorder, $W=15.0$ and $W/T=15.0$, $\beta$ stops decreasing and becomes difficult to determine due to the crossover to the exponential scaling, $t_{Th} \sim t_H \sim 2^L$, characteristic of a localized phase. Remarkably, in the Floquet case, both the spectral form factor and the power spectrum yield very similar estimates of the Thouless time for any disorder value and system size, confirming the expectation that the Floquet model exhibits less finite-size effects compared with the Hamiltonian case.
On the other hand, we observe a discrepancy at weak disorder $W\lesssim 8$ between the extracted Thouless time from the spectral form factor and the power spectrum in the Hamiltonian model in Fig.~\ref{fig:dynamical_exponent}(a)). The result from the spectral form factor seems to have a much smaller derivative with respect to $W$ between $W=4.0$ and $W=6.0$ compared to the dynamical exponent extracted from the power spectrum in the same range.
One possible cause of this discrepancy could be due to the spectral cutoff used for mitigating unwanted edge effects (see Appendix~\ref{appendix:thouless_time} for more details). Discarding $20\%$ of the states at both edges implies a reduction of the energy bandwidth. This effective bandwidth $\Delta E$ sets the smallest time scale we have access to ($\sim 1/\Delta E$). In the static Hamiltonian model the bandwidth is proportional to $W$ and independent of $L$, therefore at small $W$s and $L$s the Thouless time might be out of reach because it is smaller than the shortest accessible time scale.
When the Thouless time is getting close to the inverse bandwidth, both spectral measures will be strongly affected by edge effects even in the presence of the filter Eq.~\eqref{eq:sff_filter}. This effect can be observed in Fig.~\ref{fig:thouless_time} at $L=8$ where $t_{Th}$ is almost equal for $W=4.0$ and $W=6.0$. As a result, the spectral form factor and the power spectrum might behave differently due to edge effects yielding different scaling when the matrix size is not big enough for bringing the Thouless time within the bulk of the spectrum. The latter strongly limits the study of the Thouless time scaling at low disorder in the current system sizes. This effect is absent at the intermediate disorder because the Thouless time is larger than the inverse energy bandwidth.

\section{Conclusions}\label{sec:conclusion}
In this work, we have studied the long-range spectral correlations in the Anderson model on
the random regular graph in terms of two probes: the spectral form factor and the power spectrum.
We provide numerical evidence that the sub-diffusive phase in the Anderson model on the random regular graph, claimed in~\cite{Biroli2017delocalized,Bera2018return,DeTomasi2019subdiffusion,Biroli2020anomalous}, can be probed by the global spectral statistics which does not include any information about the eigenstate structure.
To reduce finite-size effects due to mobility edges and energy dependence of the density of states, we remove the constraint of energy conservation by introducing a Floquet version of the Anderson model on the RRG. The Floquet model has the advantage of having a structureless density of states and the absence of mobility edges.

In this setting, we extract the Thouless energy from both the spectral form factor and the power spectrum which agree with each other in the reliable range of Hamiltonian parameters and show good algebraic scaling with the diameter of RRG.
The above scaling of the extracted Thouless energy with the graph size allows us to unambiguously observe the sub-diffusive character of this dependence in the entire delocalized phase and find the exponent $\beta<1$ of the sub-diffusion.
This finding is in a good agreement with a recent random-matrix consideration~\cite{Khaymovich2021dynamical} of the RRG where the dynamical characteristics of the model, like the return probability, has been claimed to show the sub-diffusion up to zero disorder $W=0$.
As the Anderson model on the RRG provides a proxy for the dynamics of more realistic many-body disordered systems, our consideration provides the ground to the tentative sub-diffusive phase in the finite size many-body localized regime \cite{Morningstar2021avalanches} close to the true MBL transition and to its observation with the global spectral probes like the spectral form factor~\cite{Suntajs2020quantum,Sierant2020thouless,Abanin2021distinguishing}.

\section{Acknowledgments}
I.~M.~K. and G.~D.~T. acknowledge S.~Bera, V.~E.~Kravtsov, and A.~Scardicchio for works on related topics.
The work of I.~M.~K. is supported by Russian Science Foundation (Grant No. 21-72-10161). G.~D.~T. acknowledges the support from the EPiQS Program of the Gordon and Betty Moore Foundation.
D.~J.~L. acknowledges support by the DFG through SFB 1143 (project-id 247310070) and the cluster of excellence ML4Q (EXC2004, project-id 390534769).

\appendix
\section{Unfolding of the Hamiltonian spectrum}\label{appendix:unfolding}

\begin{figure}[tbh]
    \centering
    \includegraphics[width=0.9\columnwidth]{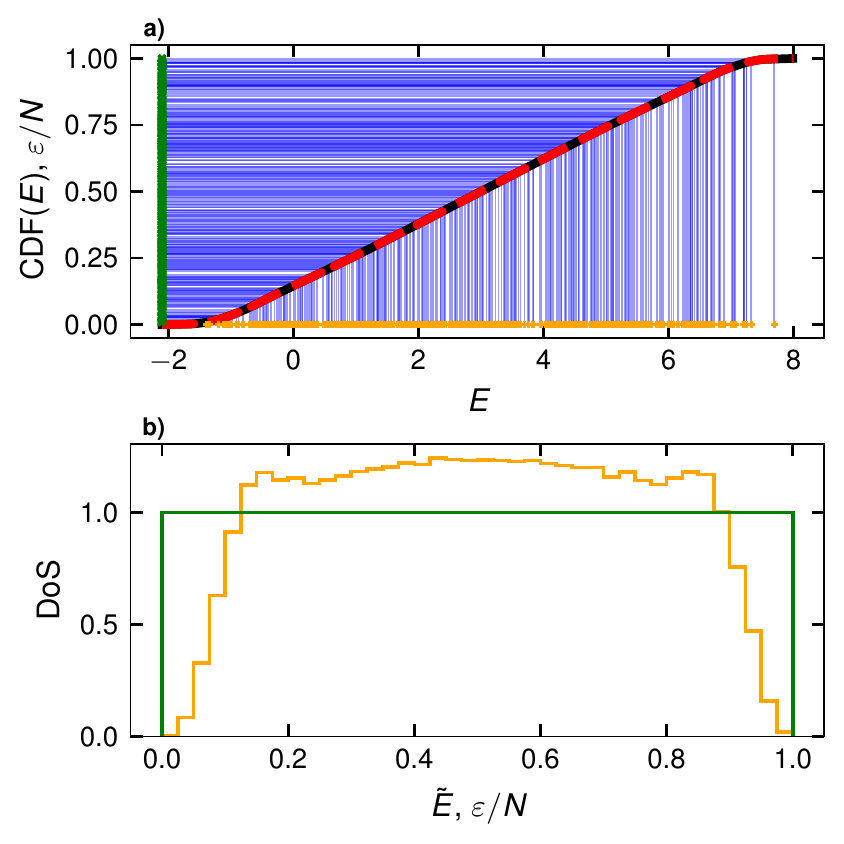}
    \caption{Unfolding of Hamiltonian spectra with parameters $L=8$ and $W=6.0$. (a)~average cumulative distribution function (CDF) as a function of eigenvalues $E$ (black) and its fitting using cubic spline (dashed red). Orange points are $E$ values of a single realization, green points are their corresponding unfolded values $\varepsilon/N=$CDF$(E)$. (b)~Density of states of the unfolded (folded) eigenvalues $\varepsilon$ ($\tilde{E}$) average over 1000 graph and diagonal disorder realizations. $\tilde{E}=(E-E_\text{min})/(E_\text{max}-E_\text{min})$ is the folded spectrum shifted and rescaled to be between 0 and 1.  }
    \label{fig:unfolding}
\end{figure}

In this section the unfolding procedure sketched in Sec.~\ref{sec:unfolding} is further explained. This procedure is illustrated in Fig.~\ref{fig:unfolding}(a). The black line corresponds to the CDF based on  $n_\text{real}=1000$ realizations of spectra of Hamiltonian~\eqref{eq:model}, with $W=6.0$ and $L=8$. The red dashed line is the averaged $\text{CDF}^\text{avg}(E)$ fitted using a cubic spline. The orange points are the eigenvalues $E^{(i)}$ of a single realization of the Hamiltonian model whilst the green points are their unfolded images $\varepsilon^{(i)}/N = \text{CDF}^\text{avg}(E^{(i)})$. In the lower panel the density of states of both folded and unfolded spectra are shown. By design, the density of states of the unfolded spectrum is constant. The folded spectra are rescaled to the interval [0,1] for perspective purposes.

\section{Extracting Thouless time from spectral form factor and power spectrum}\label{appendix:thouless_time}
\begin{figure}[t!]
	\includegraphics[width=0.9\columnwidth]{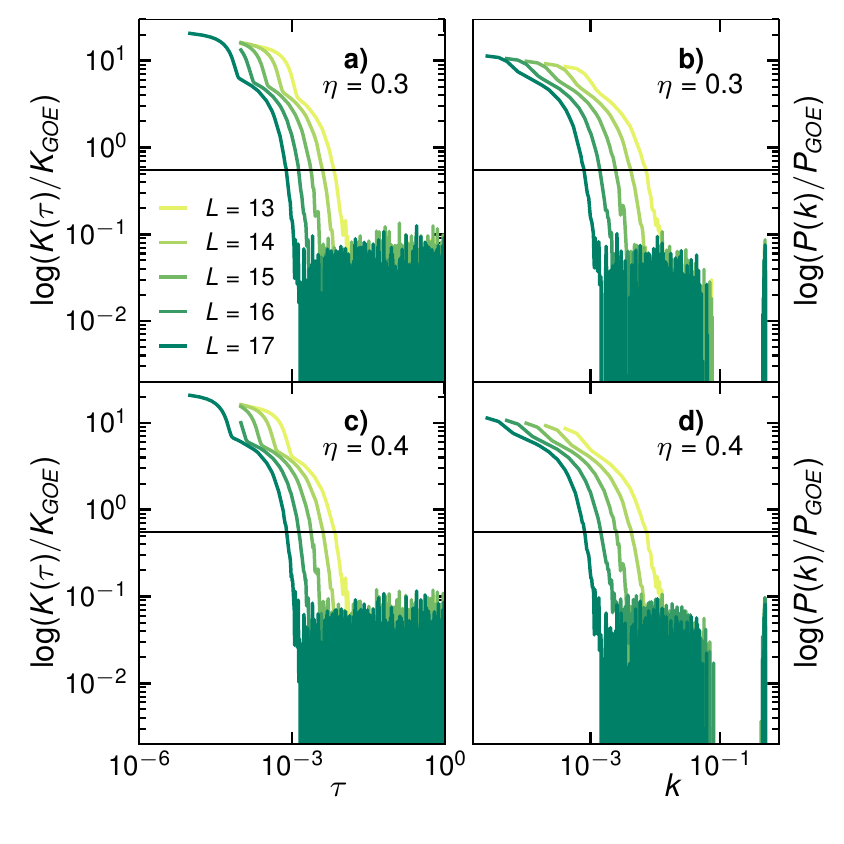}
\caption{Measures to estimate the Thouless time in the Hamiltonian model: $\log (F/F_{GOE})$ with $F$ equal to the spectral form factor (a,~c) and the power spectrum (b,~d) for the filtering parameter $\eta=0.3$ (a,~b) and $\eta=0.4$ (c,~d). Disorder is set to $W=6.0$.}
	\label{fig:appendix_hamiltonian}
\end{figure}

\begin{figure}[b!]
	\centering
	\includegraphics[width=0.9\columnwidth]{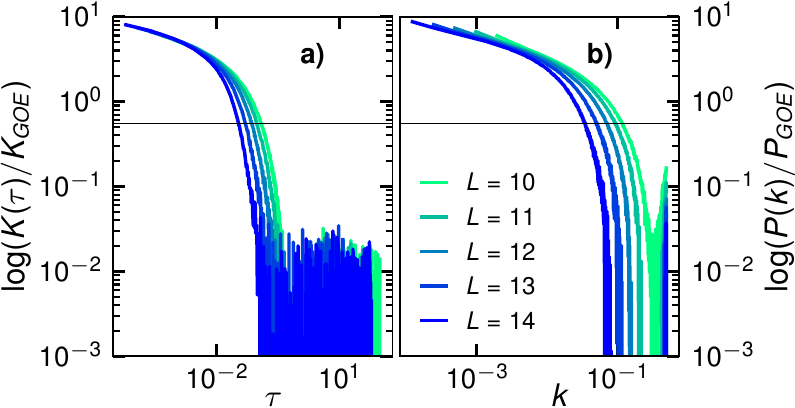}
	\caption{Measures to estimate the Thouless time in the Floquet model: $\log (F/F_{GOE})$ with $F$ equal to the spectral form factor (a) and the power spectrum (b). Disorder is set to $W/T=10.0$ .}
	\label{fig:appendix_floquet}
\end{figure}

In this section we explain the procedure of the extracting of the Thouless time from the equidistant spectra with unit mean level spacing $\varepsilon \in$ [0,$N$]. In the Floquet case (when spectral edge effects are not present), the spectral form factor and the power spectrum are computed without filtering and then compared to the GOE functions $K_\text{GOE}(\tau)=2\tau$~\cite{Brezin1997spectral} and $P_{\text{GOE}}(k)=N/2\pi^2 k$~\cite{faleiro2004theoretical} mentioned in sections~\ref{sec:model_sff} and~\ref{sec:model_ps}, respectively. Henceforth we refer only to the spectral form factor, however the procedure is the same for both quantities with the caveat that $k/N$ is used instead of $k$. At early $\tau$ we have $K>K_\text{GOE}$, consequently we identify the Thouless time as the smallest $\tau$ for which  $K=K_\text{GOE}$, we instead use $\log (K/K_\text{GOE})$ as a measure of the distance between the two functions. In practice, it is useful to set a threshold like $\log ( K(\tau_\text{Th})/K_\text{GOE} )=0.4$ and take the time that satisfies this relation as $\tau_\text{Th}$. This is illustrated in Fig.~\ref{fig:appendix_hamiltonian} and~\ref{fig:appendix_floquet}. It can be seen that a finite but not too small threshold captures the system size scaling of the curves which encodes the Thouless time system-size scaling as well. We have verified that our results are not affected by the choice of the threshold beyond the noise fluctuations, see Figs.~\ref{fig:appendix_hamiltonian} and~\ref{fig:appendix_floquet}.

In the Hamiltonian case, states at the edges of the spectrum are usually more localized and so have strong unwanted influence on the spectral measures. In order to remove such spurious effects, the spectrum is first cut off from the edges. Throughout this work 20\% of the states at each edge have been discarded leaving 60\% of the total spectrum centered at the middle of it. At the end, one works with a spectrum of reduced dimension $\tilde{N} = \lfloor 0.6 N \rfloor$. This cut is made after the unfolding, meaning that the unfolding is carried out on the whole spectrum before cutting off the edges.
Edge effects are further mitigated by multiplying a Gaussian function centered at the middle of the spectrum and reduced variance compare to the effective spectral width. In other words one does $\exp(-i2\pi\varepsilon\tau)\rightarrow g(\varepsilon)\exp(-i2\pi\varepsilon\tau)$ with $g(\varepsilon)$ given in Eq.~\eqref{eq:sff_filter}. The same applies to the spectral statistic $\delta_n=\varepsilon_n-n$ where its Fourier transform carries a weight $g(\varepsilon_n)$ given by Eq. (\ref{eq:ps_filter}).
The parameter $\eta$ controls the variance of the Gaussian filter, in Fig.~\ref{fig:appendix_hamiltonian} the function $\log( K/K_\text{GOE})$ is shown with $\eta=0.3$ and $\eta=0.4$. The small difference in filter width does not have major effects. We have set $\eta=0.3$ throughout the main text.  Next, the spectral form factor, the power spectrum, and Thouless time are computed as explained above.

\begin{figure}[tbh]
	\centering
	\includegraphics[width=0.8\columnwidth]{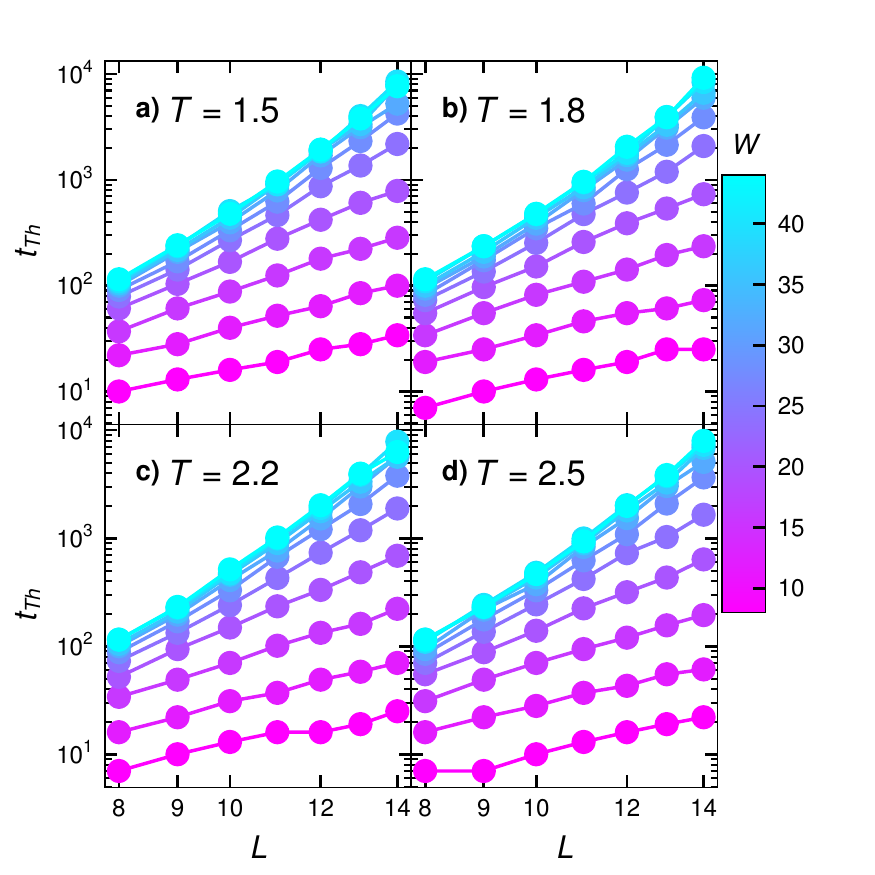}
		\caption{Thouless time computed from the spectral form factor at periods $T=1.5$, $1.8$, $2.2$, $2.5$ and disorder $W=8$, $12$, $16$, $20$, $24$, $28$, $32$, $36$, $40$, $44$. The procedure for extracting the Thouless time is the same explained in Fig.~\ref{fig:thouless_time} of the main text.}
	\label{fig:appendix_thouless_period}
\end{figure}

\begin{figure}[tbh]
	\centering
	\includegraphics[width=0.8\columnwidth]{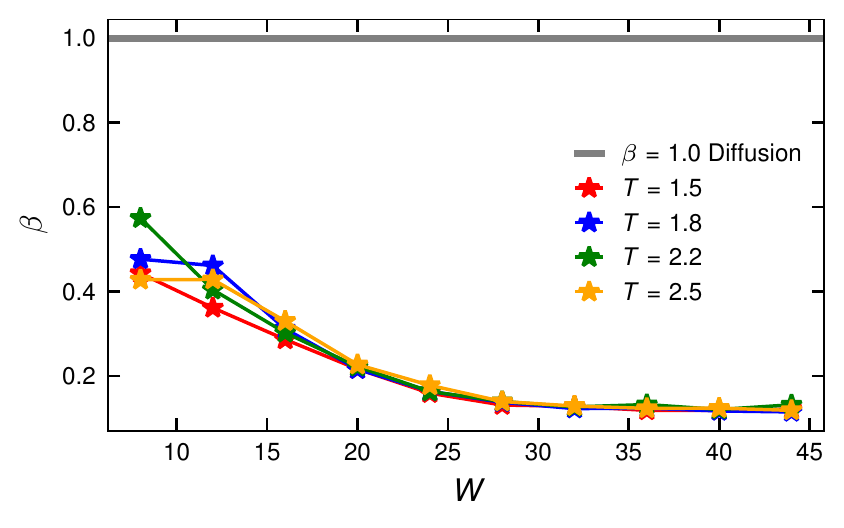}
	\caption{Dynamical exponent $\beta$ extracted from curves in Fig.~\ref{fig:appendix_thouless_period}. The same fitting procedure explained in caption of Fig.~\ref{fig:dynamical_exponent} of the main text was used. }
	\label{fig:appendix_dynamical_exponent_period}
\end{figure}

\section{Sub-diffusion at different driving period}\label{appendix:different_periods}

In order to check the sensitivity of sub-diffusion in the Floquet-driven RRG to changes in the half driving period $T$ we have computed the spectral form factor for $T=1.5$, $1.8$, $2.2$, $2.5$. Recall that in the main text $T=2.0$ is fixed. First of all, the gap ratio shows the crossing of the curves for different system sizes at roughly the same disorder value for all $T$ (not shown), therefore we still see a localization transition within the same range of disorder. In Fig.~\ref{fig:appendix_thouless_period} the Thouless time extracted from the spectral form factor is plotted. We can see that at low disorder the scaling the Thouless time is polynomial in $L$ regardless of the half period $T$. Indeed Fig.~\ref{fig:appendix_thouless_period} is quite similar to panels (c) and (d) of Fig.~\ref{fig:thouless_time}.

In order to confirm a sub-diffusive character of the transport for different $T$, we estimate the dynamical exponent fitting $t_{Th}\sim L^{1/\beta}$ for each curve in Fig.~\ref{fig:appendix_thouless_period}. The resulting dynamical exponent $\beta$ is shown in Fig.~\ref{fig:appendix_dynamical_exponent_period} as a function of disorder strength $W$. We can see that around $W\gtrsim 25$ the dynamical exponents saturate pointing out the localization threshold for all periods. At weak disorder strengths $\beta<1$ independently of $T$. The latter suggests that the sub-diffusion regime does not arise due to fine-tuned choice of driving period in the model of Eq.~\ref{eq:model_floquet}, but it is quite robust and generic.

\bibliography{references}

\end{document}